# Content and Engagement Trends in COVID-19 YouTube Videos: Evidence from the Late Pandemic

Nirmalya Thakur
Department of Electrical Engineering and Computer Science
South Dakota School of Mines and Technology
Rapid City, SD 57701, USA
nirmalya.thakur@sdsmt.edu

Madeline D Hartel
Department of Electrical Engineering and Computer Science
South Dakota School of Mines and Technology
Rapid City, SD 57701, USA
madeline.hartel@mines.sdsmt.edu

Lane Michael Boden
Department of Mechanical Engineering
South Dakota School of Mines and Technology
Rapid City, SD 57701, USA
lane.boden@mines.sdsmt.edu

Dallas Enriquez
Department of Electrical Engineering and Computer Science
South Dakota School of Mines and Technology
Rapid City, SD 57701, USA
dallas.enriquez@mines.sdsmt.edu

Boston Joyner Ricks
Department of Electrical Engineering and Computer Science
South Dakota School of Mines and Technology
Rapid City, SD 57701, USA
boston.ricks@mines.sdsmt.edu

*Abstract*—This work investigated about 10,000 COVID-19-related YouTube videos published between January 2023 and October 2024 to evaluate how temporal, lexical, linguistic, and structural factors influenced engagement during the late pandemic period. Publishing activity showed consistent weekday effects: in the first window, average views peaked on Mondays at 92,658; in the second, on Wednesdays at 115,479; and in the third, on Fridays at 84,874, reflecting a shift in audience attention toward mid- and late week. Lexical analysis of video titles revealed recurring high-frequency keywords related to COVID-19 and YouTube features, including *COVID*, *coronavirus*, *shorts*, and *live*. Frequency analysis revealed sharp spikes, with *COVID* appearing in 799 video titles in August 2024, while engagement analysis showed that videos titled with *shorts* attracted very high views, peaking at 2.16 million average views per video in June 2023. Analysis of sentiment of video descriptions in English showed weak correlation with views in the raw data (Pearson r = 0.0154, p = 0.2987), but stronger correlations emerged once outliers were addressed, with Spearman r = 0.110 (p < 0.001) and Pearson r = 0.0925 (p < 0.001) after IQR-based exclusion. The number of videos in English increased in later windows, reaching 2,440 in the final period, while videos in Hindi and Arabic continued to emphasize different categories, with news and politics being dominant in English and Hindi, and people and blogs being more prevalent in Arabic. Category-level analysis of video durations revealed contrasting outcomes: long videos focusing on people and blogs averaged 209,114 views, short entertainment videos averaged 288,675 views, and medium-to-long news and politics videos averaged 51,309 and 59,226 views, respectively. These results demonstrate that engagement patterns of COVID-19-related videos on YouTube during the late pandemic followed distinct characteristics driven by publishing schedules, title vocabulary, topics, and genre-specific duration effects. These results highlight that late-pandemic YouTube engagement followed systematic temporal, lexical, and structural patterns that can inform the design of more effective digital health communication during future public health crises.

## I. INTRODUCTION

The COVID-19 pandemic resulted in one of the most significant shifts in information ecosystems in recent history [1-3]. From the beginning of the outbreak, social media platforms have been among the primary sources of generating and disseminating information related to the origin of the virus, preventive measures, vaccination, and treatment [4-6]. Due to its global reach, YouTube has played a significant role in the public discourse about COVID-19 since the beginning of the outbreak [7]. As of July 2025, YouTube had 2.5 billion monthly users [8]. Globally, YouTube is the second most visited website after google.com [9]. It is available in 100 countries and 80 languages, with users collectively watching about 5 billion videos daily [10].

The continued reliance on YouTube for health information persists, even several years after the initial COVID-19 outbreak. This underscores the need for systematic investigations into its operation as a health communication medium. The role of YouTube as a source of both credible health information and misinformation related to COVID-19 has been the subject of ongoing discussions and debate in the scientific community [11-16]. A letter signed by more than 80 groups, including Full Fact in the UK and the Washington Post's Fact Checker, highlighted the presence of misinformation about COVID-19 and false narratives regarding the United States presidential election on the platform [17]. The letter urged YouTube to commit to funding independent research into misinformation campaigns on the platform, provide links to rebuttals within videos that distribute misinformation, cease promoting repeat offenders through its algorithm, and increase efforts to tackle misinformation in videos that are published in languages other than English [17].

Although concerns about misinformation remain part of the broader discussion, a more fundamental question is how YouTube structures the visibility and engagement of COVID-19 content in the late-pandemic period [18,19]. Several years after the outbreak, the patterns of content creation and audience engagement have undergone substantial shifts [20]. Viewers now engage with information under conditions of long-term management rather than acute crisis, and creators have adapted to new YouTube features, such as Shorts and live streaming [21-23]. The continued presence of COVID-19 content across multiple categories, from news and education to lifestyle and entertainment, demonstrates that YouTube has become more than a source of information exchange; it has become a long-term record of how health narratives are created, sustained, and amplified [24].

Prior works in this field have been typically confined to the early stages of the pandemic and have had a limited scope. They relied heavily on keyword-based searches or a subset of the most-viewed videos, which may bias the results toward outliers and limit the scope for investigating the central tendencies of engagement. Most of these works also focused primarily on videos in English, despite YouTube's multilingual reach. As a result of these limitations, several critical research questions related to videos on COVID-19 on YouTube remain unanswered [25-27] as outlined below:

1. How does the time of publication of a video influence engagement in a late-pandemic environment?
2. Which lexical features in video titles correspond to attention cycles, and how do these features align with patterns of audience engagement?
3. To what extent does sentiment in video descriptions relate to visibility when engagement may be influenced disproportionately by a small number of outlier videos?
4. How do language and content categories interact over time, and how does the duration of a video determine audience reach across genres?

The work presented in this paper addresses these research gaps through a comprehensive analysis of approximately 10,000 YouTube videos related to COVID-19, published during the late stages of the pandemic. The investigation involved analyzing publishing activity across late-pandemic windows, applying multilingual processing of video titles to identify recurring lexical patterns, evaluating sentiment in video descriptions using VADER (Valence Aware Dictionary and sEntiment Reasoner), examining linguistic and categorical variation to identify dominant themes, and assessing video duration as a factor influencing engagement across genres. Beyond advancing knowledge in this field of research, this work offers practical applications. Public health agencies can utilize the findings on temporal patterns of engagement to align video publishing schedules with periods of increased engagement on YouTube. Media organizations and educators can use the results of lexical and categorical analyses to refine their video content for diverse linguistic communities, thereby expanding their reach across different audience groups. Insights on duration effects can drive decisions about how video length interacts with topical focus and category, enabling content creators to choose video lengths that are most effective for specific themes. More broadly, the work presented in this paper demonstrates how systematic analysis of YouTube video data can be leveraged to monitor, interpret, and strategically adapt health communication in response to future public health challenges.

The rest of this paper is organized as follows. Section II presents a review of recent research works in this field and discusses the research gaps in detail. The methodology is explained in Section III, which is followed by the results in Section IV. Section V concludes the paper, which is followed by the references.

## II. LITERATURE REVIEW

Multiple prior works in this field have investigated the role of social media platforms related to the dissemination of information and misinformation about COVID-19 [28-34]. Researchers have approached this topic from multiple perspectives, including the accuracy and reliability of medical content, the dissemination of vaccine-related narratives, and the use of video for education and community building, just to name a few. These studies demonstrate the influence of social media platforms during the pandemic, while also highlighting recurring concerns regarding content quality, accessibility, and the persistence of misinformation. The rest of this section reviews works that specifically focused on YouTube, outlining the diverse methodological approaches and findings that have shaped the current understanding of YouTube's role in COVID-19 discourse.

Lee et al. [35] studied the 50 most-viewed YouTube videos that focused on the effect of vitamin C on COVID-19. Their work focused on investigating the reliability, quality, and accuracy of these videos. Out of these videos, they found that 54% were unreliable, 62% had poor quality, and 74% lacked accurate information. McDonough et al. [36] investigated the impact of YouTube-delivered intervention sessions on sedentary behavior, physical activity, and sleep quality over 12 weeks during the COVID-19 pandemic. The findings showed that these sessions improved muscle strength and sleep quality during the pandemic. Ginossar et al. [37] studied YouTube videos shared in about 144 million COVID-19 vaccine-related tweets, published between February 1, 2020, and June 23, 2020. These tweets referenced 2,097 unique YouTube videos on COVID-19. The authors used Latent Dirichlet Allocation topic modeling and independent coding to infer that conspiracy theories dominated in these YouTube videos. Sui et al. [38] explored patterns in the likes, comments, and views of 10 YouTube channels that posted home fitness videos during COVID-19 from March 2020 to June 2020. The findings suggested that the number of views on these videos declined by 24,700 per day on average. The study also showed that channels with more subscribers experienced a higher rate ($p < 0.04$) of decline in views, likes, and comments on such videos compared to channels with fewer subscribers.

Fraserv et al. [39] studied music performances broadcast via YouTube to understand the role of music in social cohesion, intercultural understanding, and community resilience during COVID-19. The studied videos represented virtual choirs, orchestras, and collaborations of various genres, including classical, pop, and fusion styles from different cultures. They identified five themes - interaction, unity, resilience, identity, and emotion in the comments of these YouTube videos. Gruzd et al. [40] studied the role of Facebook and YouTube in the creation and dissemination of COVID-19 vaccine-related misinformation. To understand the level of exposure, the authors applied a unidirectional information-sharing model. Using this approach, they studied interaction patterns on Facebook and YouTube that specifically focused on Facebook users observing a COVID-19 vaccine-related YouTube video, following this video on YouTube, and getting exposed to a list of similar videos recommended by YouTube. The results showed that despite efforts by Facebook and YouTube, people were exposed to COVID-19-related misinformation on these social media platforms.

Memioglu et al. [41] studied the content, reliability, and quality of YouTube videos focusing on the association of myocarditis with COVID-19. The authors included the upload time, video length, type of image, content, quality, the number of daily and total views, likes, dislikes, comments, and VPI of the 50 most viewed videos for their analysis. They found that the mean duration of these videos was $6.25 \pm 5.20$ minutes. Their work showed that these videos focused on general information, COVID-19, vaccination, diagnosis, patient experience, and treatment. This work also showed that the uploaders of these videos included physicians, hospital channels, health channels, patients, and others, with the most viewed, liked, and disliked videos being the videos uploaded by health channels. Yüce et al. [42] used keywords such as 'COVID-19 and dental practice', 'SARS-CoV-2 and dental practice', and '2019-COV-2 and dental practice' to collect YouTube videos about COVID-19 for analysis. The authors included 55 videos from the search results in their study. The findings showed that only 3.6% of the videos were found to be of good quality, and 43.6% of poor quality. Elareshi et al. [43] used Structural Equation Modeling, Artificial Neural Network, and importance-performance map analysis to investigate the impact of YouTube video content on Jordanian university students' behavioral intention regarding e-learning acceptance. The findings indicated that the behavioral intention of users to adopt e-learning was significantly affected by their performance expectancy and effort expectancy. This work also showed that the respondents were willing to adopt new e-learning technologies for their education.

Sadiq et al. [44] used the Theory of Planned Behavior to analyze the tone of Nigerian YouTube video titles and the tone of YouTube users' comments to investigate the causes of COVID-19 vaccine hesitancy. The authors studied YouTube videos uploaded between March 2021 and December 2022. The findings showed that the percentage of videos that expressed a positive, negative, and neutral tone was 53.5%, 40.5%, and 6%, respectively. The findings also showed that 62.6% of the comments on these videos were neutral, 32.4% were negative, and 5% were positive. This work also indicated that people's lack of trust in the government and the presence of vaccine conspiracy theories were the leading causes of COVID-19 vaccine hesitancy in Nigeria.

In summary, the following research gaps still exist in this field:

1. Most prior works focused on YouTube videos published during the first few months of the COVID-19 outbreak and did not investigate how YouTube engagement patterns evolved during the late-pandemic stage.
2. Many studies included YouTube videos that represented specific topics (such as physical activity, home fitness, vaccines, and music). The findings of such a narrow focus may not be generalizable to all YouTube videos related to COVID-19.
3. Even though some works used advanced Natural Language Processing techniques, many relied on manual coding and frequency analysis. Furthermore, no prior work analyzed lexical patterns, sentiment in descriptions, or category dynamics across different time windows.
4. Most of these works focus on videos where the language was English, despite YouTube's global scale. The findings of such studies may not apply to videos published in languages other than English.
5. Even though a few studies reported average video length, those works did not investigate how the duration of a video influences engagement. This leaves a critical gap, given the rise of YouTube Shorts and live streaming in the last few years.
6. None of the prior works investigated the effect of upload time, language, sentiment, category, and video duration in driving engagement on YouTube.

The work presented in this paper aims to address these research gaps. The step-by-step methodology that was followed in this work is presented in Section III.

### III. METHODOLOGY

The YouTube dataset used for this work was presented in a recent paper by Su and Thakur [45]. This dataset contains 9,325 YouTube videos about COVID-19 published on YouTube between January 1, 2023, and October 25, 2024. This dataset was developed by writing a program in Python 3.10 that performed data mining by connecting to the YouTube API. This program used keyword search for the data mining process, and the keywords that were used to collect COVID-19-related videos were "COVID," "COVID19" "coronavirus," "COVID-19", "corona," and "SARS-CoV-2". For each video that contained at least one of these keywords in the title or the description, the program stored the URL of the video, video ID, video title, video description, and date of publication of the video in a CSV file. After the data mining process was completed and duplicate results were removed, the authors used the yt_dlp package in Python [46] to extract the number of views, likes, comments, duration (in seconds), categories, tags, language, and caption availability per video and stored the results as separate attributes in the dataset.

Multiple programs were written in Python 3.10 to perform the investigation presented in this paper. First of all, the date of publication of each video was assigned to one of three non-overlapping time periods: January 1, 2023, to July 31, 2023, August 1, 2023, to February 29, 2024, and March 1, 2024, to October 25, 2024, to facilitate comparisons over time. The date of publication for each video was analyzed to infer the day of the week (Monday through Sunday) on which each video was published. For every day of the week, the average of views, likes, and comments was computed. To reduce sensitivity to outliers, medians and winsorized means (i.e., clipping the lowest and highest one percent of the observed distribution before averaging) were computed alongside the means. The lexical analysis of titles followed a multilingual normalization pipeline, where data preprocessing included canonical text normalization to convert letters into standard forms, converting strings to lowercase, and tokenization restricted to contiguous letter sequences. A multilingual stopwords list was then developed using Python's NLTK [47] and updated with manual inspection to remove stopwords. From the resulting vocabulary, the ten most frequent tokens were identified; for this token set, two monthly series were analyzed for each window: (i) counts of video titles containing the token and (ii) mean views among videos whose titles contain the token in that month. This ensured that the frequency- and view-based series could be compared for further analysis.

To perform sentiment analysis of video descriptions (in English), VADER (Valence Aware Dictionary for Sentiment Reasoning) was used [48]. VADER is an unsupervised learning model for sentiment analysis. It uses a sentiment vocabulary based on valence scores, and its working involves a concise, rule-based framework, enabling the development of a customized sentiment analysis engine specifically designed for the language commonly used on social media platforms. VADER is available as a Python package, and its usage does not require any subscriptions. VADER was specifically used for this work as it is a highly accurate unsupervised learning model for sentiment analysis and has been used in several prior works in this field [49-54]. The video descriptions were normalized by removing URLs, user mentions, and non-alphabetic characters. The sentiment of these descriptions was quantified using the compound sentiment score from VADER, mapping the sentiment of each video description to [−1, 1]. The association between sentiment and views was then evaluated using multiple approaches: Pearson product–moment correlation with exact p-values on the raw pairing; Pearson correlation [55,56] after winsorizing views at the first and ninety-ninth percentiles; Pearson correlation after excluding observations flagged as outliers by the interquartile rule (multiplier 1.5) [57]; and Spearman rank correlation [58] as a nonparametric check on monotonic association. The pseudocode of the Python program, which performed sentiment analysis and different types of correlation analyses, including outlier handling, is shown in Algorithm 1.

| **Algorithm 1**: Sentiment Analysis and Correlation Evaluation |
|---|
| **Input:**<br>DF = {(d_i, v_i, ℓ_i)} for i = 1..N   // description d_i, views v_i, language ℓ_i<br><br>**Parameters:**<br>  τ_pos = 0.05, τ_neg = −0.05     // VADER compound thresholds<br>  q_L = 0.01,  q_U = 0.99      // winsor quantiles<br>  k    = 1.5              // IQR multiplier<br><br>**Output:**<br>  S  // VADER compound scores in [−1, 1] (English-only)<br>  Y  // sentiment labels in {Positive, Neutral, Negative}, length n<br>  O  // boolean outlier flags for views (IQR rule), length n<br>  CORR_SUMMARY  // table over {Pearson_raw, winsorized, IQR-trim, Spearman, log}<br>  Fig_Log  // scatter plot<br><br>**Procedure:**<br>1. I_en ← { i : ("Language" ∈ Columns(DF) AND ℓ_i indicates English) OR (detect_Language(d_i) = "en") }<br>   D ← (d_i) for i ∈ I_en;  V ← (v_i) for i ∈ I_en;  n ← |I_en|<br>2. Define phi(t):<br>     t ← Lowercase(t)<br>     t ← Remove_Urls(t)<br>     t ← Remove_Mentions_And_Hashtags(t)<br>     t ← Keep_Letters_And_Whitespace(t)<br>     t ← Collapse_Whitespace(t)<br>     return t<br>   T ← (phi(D_i)) for i = 1..n<br>3. psi ← Vader_Compound<br>   S ← ( s_i = psi(T_i) ) for i = 1..n<br>   Y_i ←<br>     "Positive" if s_i ≥ τ_pos<br>     "Negative" if s_i ≤ τ_neg<br>     "Neutral"  otherwise<br>4. m_i ← 1 if (s_i not NA) AND (V_i not NA), else 0<br>5. Q1 ← Quantile(V, 0.25); Q3 ← Quantile(V, 0.75); IQR ← Q3 − Q1<br>   LO ← Q1 − k·IQR;  HI ← Q3 + k·IQR<br>   O_i ← 1 if (V_i < LO) OR (V_i > HI), else 0<br>6. v_L ← Quantile(V, q_L); v_U ← Quantile(V, q_U)<br>   V_w_i ← CLIP(V_i, v_L, v_U)  for all i<br>7. V_log_i ← log10(V_i + 1)   for all i<br>8. (r_pr, p_pr) ← Pearson(S, V )        // raw<br>   (r_pw, p_pw) ← Pearson(S, V_w )       // winsorized<br>   (r_pi, p_pi) ← Pearson(S[O=0], V[O=0] )  // IQR<br>   (rho_sr, p_sr) ← Spearman( S, V )      // rank-based<br>   (r_pl, p_pl) ← Pearson(S, V_log )      // log-scale<br>9. CORR_SUMMARY ← Table(Method ∈ {Pearson_raw, winsorized, IQR-trim, Spearman, log},<br>         r ∈ {r_pr, r_pw, r_pi, rho_sr, r_pl},<br>         p ∈ {p_pr, p_pw, p_pi, p_sr, p_pl},<br>         N_used ∈ {len(V), len(V_w), len(V[O=0]), len(V), len(V_log) })<br>10. Fig_Log ← Scatter(x = S, y = V_log, title = "Sentiment vs log10(Views+1)")<br>11. return ( S, Y, O, Corr_Summary, Fig_Log) |

The distribution of languages and categories was then analyzed for each time window, i.e., January 1, 2023, to July 31, 2023, August 1, 2023, to February 29, 2024, and March 1, 2024, to October 25, 2024. For every window, the ten most frequently used languages, based on video count, were identified. Within each of these languages, category distributions were analyzed and ranked in the same window to identify which themes were most prominent across different languages and time periods. Thereafter, the duration of the videos was converted to a categorical attribute. The values in seconds were converted into Short (0–299), Medium (300–1,200), and Long (>=1,201) to assign these categorical labels. For every category and duration band, sample size and mean views, likes, and comments were computed. These findings provided insights into the interaction between content length, topical focus, and audience engagement.

## IV. RESULTS AND DISCUSSIONS

The analysis of views revealed distinct patterns of engagement across the three late-pandemic windows. In the first window, average views peaked on Mondays at 92,658, while remaining comparatively high on Wednesdays at 83,86.60 and on Saturdays at 59,096.73. Sundays showed lower views at 33,018.14, despite higher comments, which suggests a different mode of audience interaction late in the week. In the second window, Wednesdays and Saturdays led the views, at 115,478.65 and 98,196.41, respectively, with Tuesdays close behind at 98,228.26, indicating mid-week surges that persist even as overall conditions change. In the third window, Fridays led with 84,874.45 average views, followed by Thursdays at 74,339.32 and Mondays at 57,154.05, pointing to a late-week rise in audience activity during the final period. These findings were consistent across likes and comments, and the accompanying medians and winsorized means confirmed that the results were not driven by a small number of outlier videos. These findings are presented in Figures 1, 2, and 3, respectively.

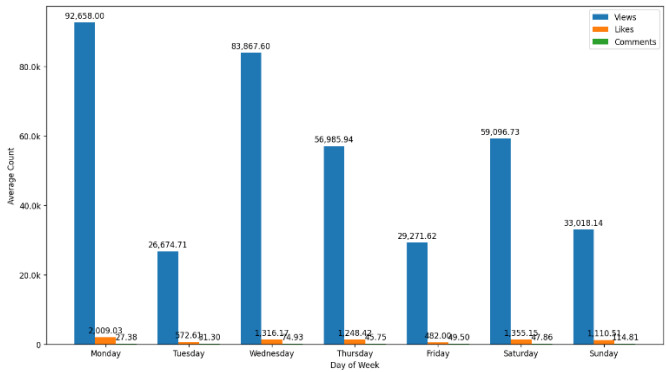

**Figure 1.** Average engagement by day of week in time window 1 (January 1, 2023, to July 31, 2023)

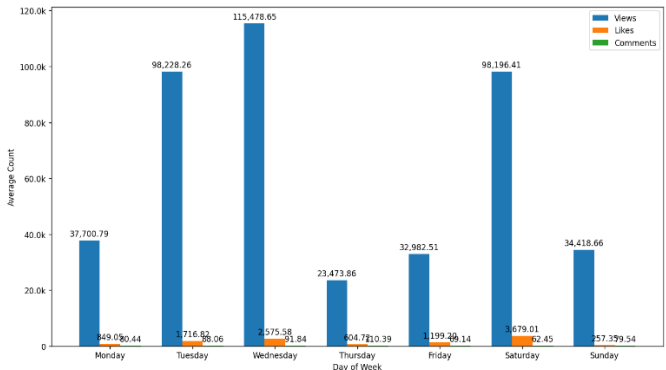

**Figure 2.** Average engagement by day of week in time window 2 (August 1, 2023, to February 29, 2024)

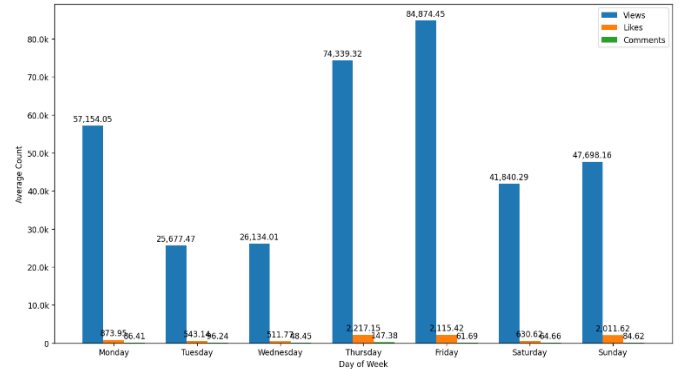

**Figure 3.** Average engagement by day of week in time window 3 (March 1, 2024, to October 25, 2024)

The findings of lexical pattern analysis in video titles revealed that the popular keywords focused on COVID-19 or YouTube features, and their usage patterns varied from month to month. These keywords included *coronavirus, COVID, news, corona, cases, koronaa, update, shorts, live,* and *India*. The analysis highlighted distinct late-period spikes, notably for *COVID*, which was present in 799 video titles in August 2024 and 716 video titles in September 2024, and for *coronavirus*, which was present in 479 video titles in September 2024. These results demonstrate that YouTube content creators started reusing keywords popular during the first few months of the outbreak to highlight recent updates and news about COVID during the late pandemic period. The analysis of average views showed that title terms were linked to very different patterns of reach. Titles containing *shorts* achieved very high average views in June 2023 at approximately 2.16 million per video, and titles with *live* were associated with very high average views in February 2023 at approximately 737,268. Videos containing *cases* showed a sharp surge in May 2024 at about 1.45 million average views, while *news* reached approximately 254,192 in May 2023, and *coronavirus* reached 200,270 in October 2023. These results are presented in Figures 4 and 5, respectively. These results indicate that recurring surges in video reach occurred when video titles contained terms related to COVID-19 or YouTube features (e.g., Shorts and live).

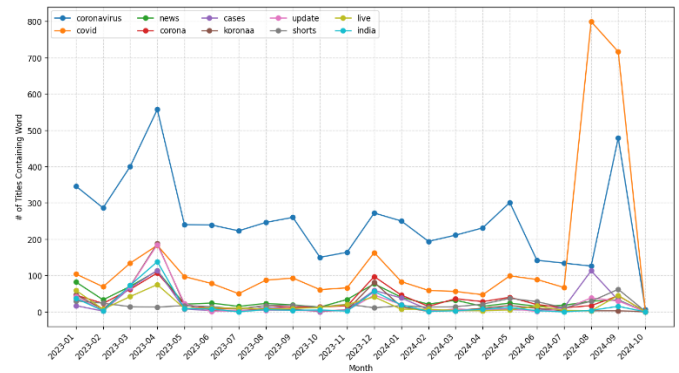

**Figure 4.** Analysis of top 10 words in video titles per month from January 1, 2023, to October 25, 2024

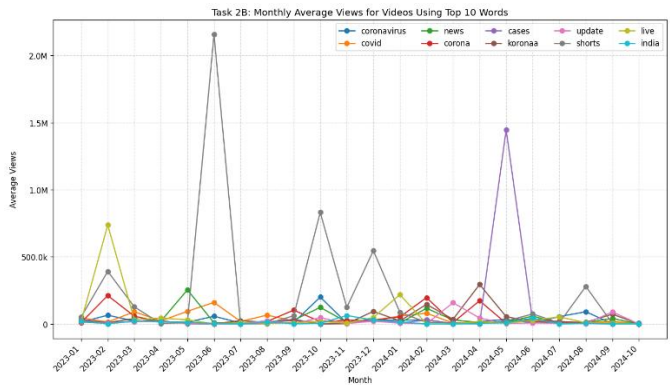

**Figure 5**. Analysis of views of video titles (containing the keywords from Figure 4) per month from January 1, 2023, to October 25, 2024

Thereafter, sentiment analysis using VADER and correlation analysis were performed. Sentiment analysis of video descriptions in English initially showed a weak correlation with views, with Pearson's r = 0.0154 and p = 0.2987. After addressing the outliers, stronger correlation patterns were observed. Winsorization at the first and ninety-ninth percentiles produced a Pearson coefficient of 0.0331 (p = 0.0256). Excluding values flagged by the interquartile rule raised the correlation coefficient to 0.0925 (p < 0.001), and the rank-based Spearman correlation on views reached 0.110 (p < 0.001). These findings show that the sentiment of video descriptions maintained a correlation with reach, and that the correlation became stronger once the influence of outliers was reduced. A log transform of views [59] vs sentiment was plotted using a scatterplot, as shown in Figure 6, for better readability of these findings. The binning of sentiment into narrow intervals was also performed, as shown in Figure 7. This analysis revealed consistent differences between medians and winsorized means, indicating that view counts followed a skewed distribution, such that medians and means captured different aspects of central tendency even after outliers were handled.

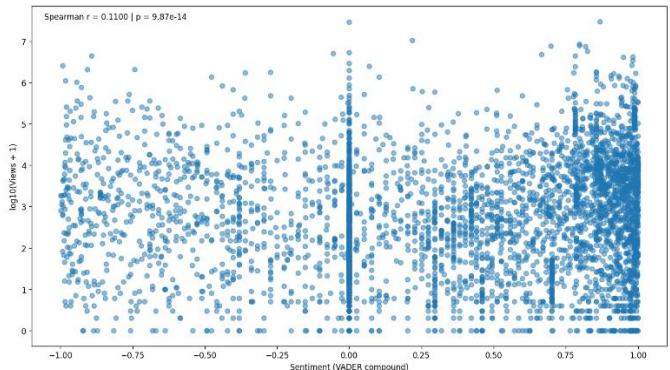

**Figure 6.** Analysis of sentiment vs log 10 (views + 1)

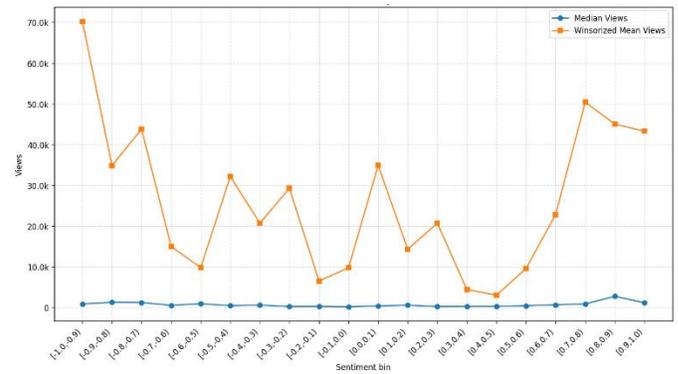

**Figure 7**. Analysis of views by sentiment bins

The analysis of patterns of language use and video categories showed noticeable variation across the three late-pandemic time windows. In the first window, English and Hindi were nearly co-dominant, with 1,118 and 1,020 videos, respectively, followed by Spanish at 206. By the third window, English had expanded substantially to 2,440 videos, while Hindi stood at 519 and Spanish at 184. These results are shown in Figures 8, 9, and 10, respectively. This distribution indicates a strong focus among YouTube content creators to publish videos in English over other languages. Within languages, category distributions showed strong focuses on news and politics for both English and Hindi in the early period, with 459 English videos and 789 Hindi videos in this category during the first window, while videos in other languages, such as Arabic, placed more emphasis on people and blogs. These results are presented in Figures 11 and 12, respectively. For paucity of space, the analysis of only English and Hindi videos from the first window is presented in this section. These findings indicate that while the share of English-language videos increased substantially, other languages maintained focus on different topics, highlighting the multilingual and evolving nature of COVID-19 discourse on YouTube during the late pandemic period.

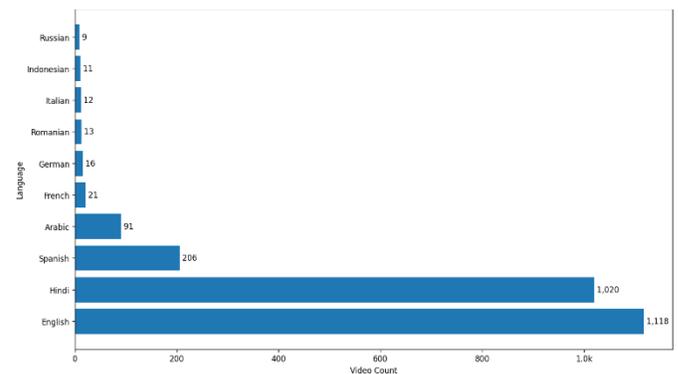

**Figure 8**. Top 10 languages by video count (window 1).

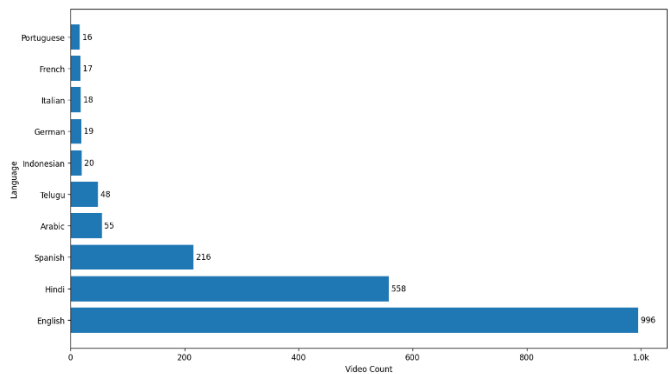

**Figure 9.** Top 10 languages by video count (window 2)

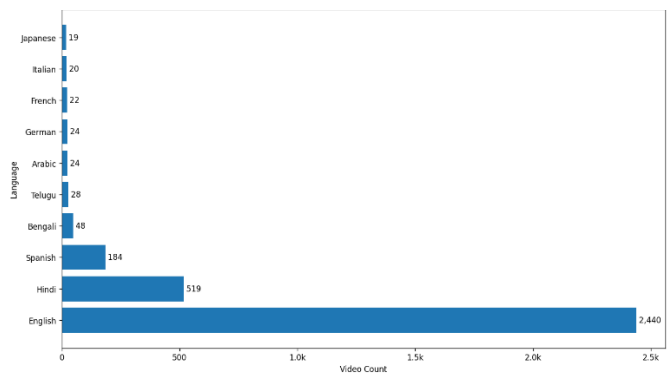

**Figure 10.** Top 10 languages by video count (window 3)

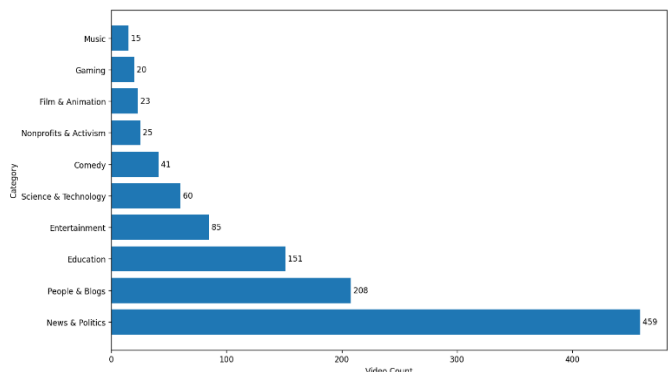

**Figure 11.** Top 10 categories the videos in English focused on in window 1.

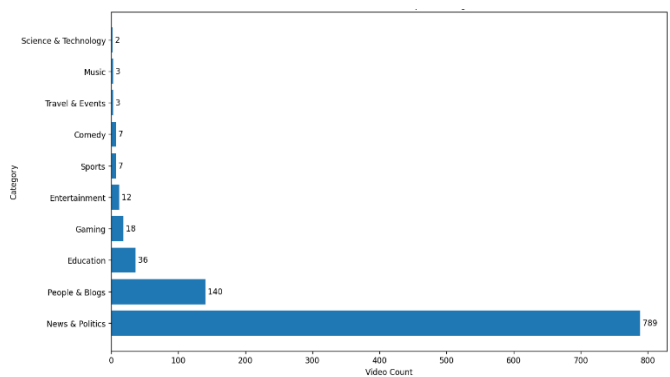

**Figure 12.** Top 10 categories the videos in Hindi focused on in window 1.

The analysis of video duration by category revealed that video duration was associated with distinct engagement patterns within each category. In the people and blogs category, long videos averaged about 209,114 views, medium videos averaged 61,468 views, and short videos averaged 33,224 views. In the entertainment category, an opposite pattern was observed, with short videos averaging about 288,675 views, exceeding both medium and long videos. Videos that focused on comedy received a very high average of views for medium-length videos, at 396,023 views. Videos that focused on news and politics showed higher averages for medium and long videos, at approximately 51,309 and 59,226 views, respectively, compared with 23,176 for short videos. These results are presented in detail in Table 1. The findings from Table 1 show that the duration-related engagement patterns of COVID-19 YouTube videos during the late pandemic period were genre and topic-specific.

In summary, this work presents a comprehensive analysis of the patterns of engagement of COVID-19 videos published on YouTube during the late pandemic period, emphasizing how temporal, linguistic, sentiment, and category factors drive engagement patterns. Weekday-level analysis showed that average reach shifted toward mid- and late-week as the period progressed. Lexical analysis demonstrated that keywords related to COVID-19 and YouTube platform features (example, Shorts and live) resulted in large-scale surges in views. Sentiment analysis revealed that raw correlations between sentiment scores of video descriptions and views were weak. However, these correlations were observed to be much stronger once outliers were addressed, with Spearman's $r = 0.110$ ($p < 0.001$) and Pearson's $r = 0.0925$ ($p < 0.001$), indicating that the sentiment in video descriptions played a significant role in shaping reach. The cross-linguistic and cross-category analysis revealed that news and politics remained central for both English and Hindi videos, while videos in other languages, such as Arabic, placed more emphasis on people and blogs. The analysis of video durations confirmed that engagement patterns varied by category: long-form content gained significant reach in people and blogs, short videos dominated in entertainment, and medium-to-long videos were more popular in news and politics, showing that duration was not a uniform driver of engagement but was based on the genre of the videos.

Several prior works in this field have emphasized that social media platforms have played a central role in shaping public understanding and influencing policy discussions since the beginning of the COVID-19 pandemic [60-65]. This highlights the importance of evidence-based strategies for content dissemination in health communication and governance. The findings of this work have several practical applications in the real world. Public health agencies and policy-making bodies can align their YouTube video publishing schedules with the mid- and late-week engagement patterns reported in this paper (Figures 1, 2, and 3), use keywords in video titles that are associated with higher views (Figures 4 and 5), and plan video durations in ways consistent with category-specific engagement patterns as shown in Table

**Table 1**. Engagement metrics (views, likes, comments) by category and video duration band (Short, Medium, Long)

| Category | Short Total | Short Avg Views | Short Avg Likes | Short Avg Comments | Medium Total | Medium Avg Views | Medium Avg Likes | Medium Avg Comments | Long Total | Long Avg Views | Long Avg Likes | Long Avg Comments |
|---|---|---|---|---|---|---|---|---|---|---|---|---|
| News & Politics | 2837 | 23176.05 | 408.07 | 40.38 | 560 | 51309.3 | 878.84 | 246.12 | 619 | 59225.86 | 575.41 | 113.39 |
| People & Blogs | 1604 | 33224.16 | 581.9 | 9.02 | 249 | 61468.34 | 860.26 | 67.77 | 255 | 209114.4 | 1778.17 | 109.03 |
| Education | 475 | 75846.42 | 1116.43 | 21.28 | 225 | 39772.26 | 2276.91 | 380.25 | 190 | 32043.66 | 1250.21 | 298.57 |
| Entertainment | 394 | 288675 | 8257.68 | 38.71 | 95 | 55587.48 | 1060.13 | 205.57 | 163 | 37493.25 | 702.88 | 73.72 |
| Science & Technology | 162 | 5783.27 | 325.17 | 16.85 | 118 | 5442.03 | 282.19 | 84.42 | 168 | 20431.52 | 364.61 | 57.4 |
| Nonprofits & Activism | 67 | 87919.15 | 3279.66 | 144.82 | 114 | 11366.68 | 420.14 | 7.2 | 79 | 234.68 | 3.23 | 0.71 |
| Film & Animation | 135 | 71183.74 | 3388.07 | 70.79 | 17 | 16378 | 219.65 | 18.35 | 9 | 496428.8 | 1489 | 39.89 |
| Music | 120 | 73567.16 | 1090.36 | 191.97 | 27 | 4890.15 | 145.41 | 6.15 | 13 | 568.62 | 14.15 | 3 |
| Gaming | 110 | 154195.5 | 4630.21 | 59.3 | 16 | 12365.12 | 497.56 | 27.44 | 20 | 49310.8 | 1914.25 | 62.7 |
| Comedy | 115 | 237273.9 | 13668.07 | 142.14 | 15 | 396023 | 13597 | 754.13 | 4 | 195929 | 3232.5 | 1614.75 |
| Howto & Style | 68 | 4728.71 | 156.66 | 4.13 | 12 | 106.67 | 6.58 | 1.67 | 10 | 27561.1 | 474.8 | 23.7 |
| Travel & Events | 58 | 16979.5 | 1028.26 | 7.86 | 13 | 2498.85 | 73.77 | 18.92 | 14 | 121 | 1.5 | 0.71 |
| Sports | 50 | 113999.3 | 4063.34 | 92.34 | 11 | 26289.36 | 529.55 | 181.09 | 8 | 236.75 | 7.38 | 1.75 |
| Pets & Animals | 28 | 1185.43 | 61.29 | 6.36 | 13 | 51924.92 | 1157.31 | 95.92 | 9 | 488.11 | 34.56 | 2.67 |
| Autos & Vehicles | 6 | 610.33 | 18.17 | 8 | 1 | 21662 | 1921 | 283 | 4 | 12 | 0 | 0 |

1. Multilingual content creators and journalists can benefit from the findings that, in the category of news and politics videos, English and Hindi were most watched (Figures 11 and 12). Educators can also apply these findings to improve the reach of their videos on YouTube. They can tailor instructional material to emphasize evidence-based strategies for audience engagement. For example, the association between video length and category-specific views (Table 1) can be incorporated into training educators who are starting a YouTube channel on how to design educational content for YouTube to maximize engagement.

This work has a few limitations. First, the dataset analyzed in this study contains the data of about 10,000 YouTube videos about COVID-19 published between January 1, 2023, and October 25, 2024. So, the results presented in this paper are based on this dataset. As YouTube is a globally popular social media platform where several videos related to different topics, containing different sentiments and languages, are uploaded every day, it is possible that if a similar data collection followed by a similar analysis (as presented in this paper) is performed, the results may vary as compared to the results presented in this paper. Second, VADER was used to perform sentiment analysis of the video descriptions published in English. Although it is a highly accurate unsupervised learning model for sentiment analysis, it is not 100% accurate. As the dataset used contains the data of about 10,000 YouTube videos, manual verification of the results from VADER was not feasible. Finally, the values of views, likes, and comments used for this study were directly used from this dataset, which hasn't been updated since October 25, 2024. Based on recent updates and news related to COVID-19, the views, likes, and comments of one or more videos in this dataset may have increased since then. So, if the most up-to-date engagement metrics for all the videos in this dataset are collected from YouTube and this study is repeated, it is possible that there may be minor variations in the findings as compared to the findings presented in this paper.

## V. CONCLUSION

This work presents a comprehensive analysis of engagement patterns in approximately 10,000 COVID-19-related YouTube videos from the late pandemic period, investigating publishing schedules, lexical choices in titles, sentiment in descriptions, language, category variation, and video duration. The results show that average reach shifted from early-week peaks to mid- and late-week highs, with Mondays averaging 92,658 views in the first window, Wednesdays rising to 115,479 in the second, and Fridays leading with 84,874 in the third. Multilingual lexical analysis of the video titles revealed recurrent surges in a set of COVID-19-specific and YouTube-specific keywords, including "covid," "coronavirus," "shorts," and "live". Notably, "covid" appeared in 799 video titles in August 2024, and "shorts" averaged 2.16 million views per video in June 2023. Sentiment analysis of video descriptions in English using VADER revealed a weak correlation with views in the raw data (Pearson $r = 0.0154$, $p = 0.2987$). However, after handling outliers, a stronger correlation emerged, with Spearman $r = 0.110$ ($p < 0.001$). Cross-linguistic analysis revealed the patterns of language usage in videos related to different categories. For example, English and Hindi videos dominated in the category of news and politics, while videos in other languages, such as Arabic, focused on people and blogs. The duration analysis of these videos showed that engagement patterns varied by genre. For example, long videos in people and blogs averaged 209,114 views, while short entertainment videos averaged 288,675, and news and politics favored medium and long formats at 51,309 and 59,226, respectively.

These findings not only advance knowledge in this field but also carry practical relevance. Public health agencies and policymakers can align their video publication schedules and utilize specific keywords associated with higher reach based on the results of this paper. They can also plan video durations in accordance with category-specific engagement patterns as discussed in this paper. Furthermore, journalists, educators, and YouTube content creators can refer to the findings of engagement patterns based on video durations for different video categories and plan their video durations to increase the reach of their videos. Future work would involve integrating the role of YouTube video recommendations, user demographics, and YouTube advertisements to determine their influence on these engagement metrics.